\documentclass[12pt]{article}
\usepackage{a4wide,epsfig,amsmath,amssymb,cite,scalefnt,graphicx}

\def\npb{{Nucl.\ Phys.\ }{\bf B}}
\def\plb{{Phys.\ Lett.\ }{ \bf B}}
\def\be{\begin{equation}}
\def\ee{\end{equation}}
\def\bea{\begin{eqnarray}}
\def\eea{\end{eqnarray}}
\def\nn{\nonumber\\}

\def\tr{\hbox{tr}}

\def\prd{{Phys.\ Rev.\ }{\bf D}}

\def\prl{Phys.\ Rev.\ Lett.\ }

\def\thetabar{{\overline\theta}}
\def\alphadot{\dot\alpha}
\def\betadot{\dot\beta}
\def\Qbar{\overline Q}
\def\psibar{\overline{\psi}}

\def\frak#1#2{{\textstyle{{#1}\over{#2}}}}

\def\ybar{\overline y}

\def\frak#1#2{{\textstyle{{#1}\over{#2}}}}

\def\lambdabar{\overline\lambda}

\def\Dtil{\tilde D}

\def\Ftil{\tilde F}

\def\sigmabar{\overline\sigma}
\def\phibar{\overline\phi}

\def\psibar{\overline\psi}
\def\Fbar{\overline F}

\def\Dtil{\tilde D}

\def\Ncal{{\cal N}}
\def\Ftil{\tilde F}
\def\alphadot{\dot\alpha}
\def\betadot{\dot\beta}

\def\pa{\partial}

\input epsf
\begin{document}

\begin{titlepage}
\begin{flushright}
LTH899\\
\end{flushright}
\date{}
\vspace*{3mm}

\begin{center}
{\Huge
Renormalisation of the  
non-anticommutativity parameter\break at two loops}\\[12mm]
{\bf I.~Jack and R.~Purdy}\\

\vspace{5mm}
Dept. of Mathematical Sciences,
University of Liverpool, Liverpool L69 3BX, UK\\

\end{center}

\vspace{3mm}
\begin{abstract}
We present evidence that the non-anticommutativity
parameter for the $\Ncal=\frak12$
supersymmetric $SU(N)\otimes U(1)$ gauge theory 
is unrenormalised through two loops.

\end{abstract}

\vfill

\end{titlepage}

\section{\label{intro}Introduction}
Deformed quantum field theories have been subject to
renewed attention in recent years due to their natural appearance
in string theory. Initial investigations focussed on theories on  
non-commutative spacetime in which the commutators of the spacetime
co-ordinates become
non-zero. More recently\cite{casal,casala,brink,schwarz,ferr,klemm,abbas,deboer,
oog}, non-anticommutative supersymmetric theories have been
constructed by deforming the anticommutators of the Grassmann co-ordinates
$\theta^{\alpha}$
(while leaving the anticommutators of the $\thetabar{}^{\alphadot}$ unaltered).
Consequently, the anticommutators of the supersymmetry generators
$\Qbar_{\alphadot}$ are
deformed 
while the remainder are unchanged. 
It can be shown that this structure arises in string theory in a background 
with a constant graviphoton field strength. A graviphoton background 
$F_{\alpha\beta}$ couples to the field $q_{\alpha}$ which is the string
worldsheet field corresponding to the supercharge $Q_{\alpha}$ (and also to its 
worldsheet conjugate) in Berkovits' formulation of the superstring\cite{berk}. 
Upon eliminating $q$ and its conjugate using their 
equations of motion, one obtains an effective contribution to the
lagrangian
\be
L_{\hbox{eff}}=\frac{1}{\alpha^{\prime 2}}
F^{-1}_{\alpha\beta}\pa\tilde\theta^{\alpha}
\pa\tilde\theta^{\beta},
\ee
where $\tilde\theta$ is the worldsheet conjugate of $\theta$.
This leads to a propagator 
\be
<\theta^{\alpha}(\tau)\theta^{\beta}(\tau')>=\alpha^{\prime2}F^{\alpha\beta}
\hbox{sign}(\tau-\tau').
\ee 
With standard open string coupling arguments, this implies
\be
\{\theta^{\alpha},\theta^{\beta}\}=\alpha^{\prime2}F^{\alpha\beta}
\equiv C^{\alpha\beta},
\ee
where $C^{\alpha\beta}$ is usually referred to as the 
``non-anticommutativity parameter''. We then find
\bea
\{\Qbar_{\alphadot},\Qbar_{\betadot}\}=&-4
C^{\alpha\beta}\sigma^{\mu}_{\alpha\alphadot}\sigma^{\mu}_{\beta\betadot}
\frac{\pa^2}{\pa y^{\mu}\pa y^{\nu}},\nn
y^{\mu}=&x^{\mu}+i\theta^{\alpha}\sigma^{\mu}_{\alpha\alphadot}
\thetabar^{\alphadot}.
\eea
(More details of this derivation can be found in 
Refs.~\cite{oog,seib}.)
It is straightforward
to construct non-anticommutative versions of ordinary supersymmetric theories
by taking the superspace action and replacing the ordinary product by
the Moyal $*$-product\cite{seib} which implements the non-anticommutativity.
Non-anticommutative versions of the Wess-Zumino model and supersymmetric gauge  
theories have been formulated in four
dimensions\cite{seib,araki} and their renormalisability
discussed\cite{brittoa,brittoaa,terash,brittob,rom,lunin},
with explicit computations up to two loops\cite{grisa} for the Wess-Zumino
model and one loop for gauge theories\cite{jjwa,jjwb,penrom,grisb,jjwc}.
Even more recently, non-anticommutative theories in two dimensions have been
constructed\cite{inami, chand, chanda, luis,chandc}, and
their one-loop divergences computed\cite{arakib,jp}.
In Ref.~\cite{jjp}
we returned to a closer examination of the non-anticommutative
Wess-Zumino model (with a superpotential) in four dimensions, and
showed that to obtain correct
results for the theory where the auxiliary fields have been eliminated,
from the corresponding results for
the uneliminated theory, it is necessary to include
in the classical action separate couplings for all the terms which may be
generated by the renormalisation process; and in Ref.~\cite{jjpa}
we extended this analysis to the gauged $U(1)$ case.

There are obstacles to obtaining
a renormalisable $\Ncal=\frak12$ theory with a trilinear superpotential 
in the case of adjoint matter (in the case of matter in the 
fundamental representation, only a mass
term is allowed anyway)\cite{jjwc}.
The requirements of $\Ncal=\frak12$ invariance and renormalisability
impose the choice of gauge group $SU(N)\otimes U(1)$ (rather than $SU(N)$ or
$U(N)$)\cite{jjwa}, \cite{jjwb}. In the adjoint case with a
trilinear superpotential, the matter fields must also be 
in a representation of $SU(N)\otimes U(1)$. The problem is that the 
superpotential 
contains terms with different combinations of $SU(N)$ and $U(1)$ chiral
fields which mix under $\Ncal=\frak12$ supersymmetry, but for which the 
Yukawa couplings renormalise differently.  
However, recently an elegant solution to this problem has been
found\cite{penroma} in which the kinetic terms for the $U(1)$ chiral fields 
are modified, in such a way that the $SU(N)$ and $U(1)$ chiral fields (and 
consequently their Yukawa couplings) renormalise in exactly the same way.
In Ref.~\cite{jjpb} we confirmed the conclusions of Ref.~\cite{penroma} in
a component version of their superspace calculation.

The results of Refs.~\cite{penroma,jjpb} imply that the non-anticommutativity
parameter ($C$)
which specifies the superspace deformation in $\Ncal=\frak12$ 
supersymmetry is unrenormalised at one loop. It is clearly interesting to ask 
whether this feature persists at higher orders. A full two-loop calculation 
would be extremely complex, and the results we present here are only partial
in two respects. Firstly we only check the renormalisation of one, 
judiciously chosen, term in the action (of course if different terms in 
the action required different renormalisations of $C$,
this would represent a violation of $\Ncal=\frak12$
supersymmetry); and secondly, we only check the 
terms in the two-loop renormalisation constant for $C$
which include the Yukawa coupling, omitting the
purely gauge-coupling dependent term. Our 
conclusion is that there are no Yukawa 
dependent terms in the renormalisation constant for $C$ through two loops, 
and we consider it likely that $C$ is unrenormalised at this order.

\section{The classical adjoint action}
In this section we present the classical form of the
adjoint $\Ncal=\frak12$ action with a superpotential in the component 
formalism, including the modifications suggested in Ref.~\cite{penroma}. 
The adjoint action was first 
introduced in Ref.~\cite{araki}\ for the gauge group 
$U(N)$. However, 
as we noted in Refs.~\cite{jjwa}, \cite{jjwb}, 
at the quantum level the $U(N)$ gauge 
invariance cannot be retained since the $SU(N)$ and $U(1)$ gauge couplings 
renormalise differently; and we are
obliged to consider a modified $\Ncal=\frak12$ invariant theory  
with the gauge group $SU(N)\otimes U(1)$. In the adjoint case with a 
Yukawa superpotential, 
it turns out that the matter fields must also be in the adjoint
representation of $SU(N)\otimes U(1)$. 
The classical action with a superpotential may be written
\bea
S_0&=&\int d^4x
\Bigl\{e^{AB}(-\frak14F^{\mu\nu A}F^B_{\mu\nu}-i\lambdabar^A\sigmabar^{\mu}
(D_{\mu}\lambda)^B+\frak12D^AD^B)\nn
&&-\frak12iC^{\mu\nu}d^{ABC}e^{AD}F^D_{\mu\nu}\lambdabar^B\lambdabar^C\nn
&&+\Fbar F-i\psibar\sigmabar^{\mu}D_{\mu}\psi-D^{\mu}\phibar D_{\mu}\phi
+\phibar D_F\phi +
i\sqrt2(\phibar \lambda_F\psi-\psibar\lambdabar{}_F\phi)\nn
&&+C^{\mu\nu}(
\sqrt2D_{\mu}\phibar\lambdabar{}_D\sigmabar_{\nu}\psi
+i\phibar F_{D\mu\nu} F)\nn
&&+
(\kappa-1)\bigl[
\Fbar^0 F^0
-i\psibar^0\sigmabar^{\mu}\pa_{\mu}\psi^0-\pa^{\mu}\phibar^0 \pa_{\mu}\phi^0\nn
&&+d^{000}C^{\mu\nu}(
\sqrt2\pa_{\mu}\phibar^0\lambdabar{}^0\sigmabar_{\nu}\psi^0
+i\phibar^0 F^0_{\mu\nu} F^0)\nn      
&&+d^{ab0}C^{\mu\nu}(\sqrt2D_{\mu}\phibar^a\lambdabar{}^b
\sigmabar_{\nu}\psi^0+i\phibar^a F^b_{\mu\nu} F^0)
\bigr]\nn
&&+\frak12\left(yd^{ABC}\phi^A\phi^B F^C
-yd^{ABC}\phi^A\psi^B\psi^C
+\ybar d^{ABC}\phibar^A\phibar^B \Fbar^C
-\ybar d^{ABC}\phibar^A\psibar^B\psibar^C\right)\nn
&&+\frak13i\ybar C^{\mu\nu}f^{abc}D_{\mu}\phibar^a D_{\nu}\phibar^b\phibar^c
-\frak{1}{3}i\ybar C^{\mu\nu}d^{ABE}d^{CDE}
F^D_{\mu\nu}\phibar^A\phibar^B\phibar^C\nn
&&+\kappa_1\sqrt2C^{\mu\nu}d^{abc}(\phibar^a\lambdabar^b\sigmabar_{\nu}
D_{\mu}\psi^c
+D_{\mu}\phibar^a\lambdabar^b\sigmabar_{\nu}\psi^c+i\phibar^aF_{\mu\nu}^b
F^c)\nn
&&+\kappa_2\sqrt2C^{\mu\nu}d^{ab0}(\phibar^0\lambdabar^a\sigmabar_{\nu}
D_{\mu}\psi^b
+\pa_{\mu}\phibar^0\lambdabar^a\sigmabar_{\nu}\psi^b
+i\phibar^0F_{\mu\nu}^aF^b)\nn
&&+\kappa_3\sqrt2C^{\mu\nu}d^{ab0}(\phibar^a\lambdabar^b\sigmabar_{\nu}
\pa_{\mu}\psi^0
+D_{\mu}\phibar^a\lambdabar^b\sigmabar_{\nu}\psi^0
+i\phibar^aF_{\mu\nu}^bF^0)\nn
&&+\kappa_4\sqrt2C^{\mu\nu}d^{0ab}(\phibar^a\lambdabar^0\sigmabar_{\nu}
D_{\mu}\psi^b
+D_{\mu}\phibar^a\lambdabar^0\sigmabar_{\nu}\psi^b
+i\phibar^aF_{\mu\nu}^0
F^b)\nn
&&+\kappa_5\sqrt2C^{\mu\nu}d^{000}(\phibar^0\lambdabar^0\sigmabar_{\nu}
\pa_{\mu}\psi^0
+\pa_{\mu}\phibar^0\lambdabar^0\sigmabar_{\nu}\psi^0
+i\phibar^0F_{\mu\nu}^0
F^0)\Bigr\}.
\label{Sadj}
\eea
where
\bea
\lambda_F&=&\lambda^A\Ftil^A,\quad (\Ftil^A)^{BC}=if^{BAC},\nn
\lambda_D&=&\lambda^A \Dtil^A,\quad (\Dtil^A)^{BC}=d^{ABC},
\label{lamdef}
\eea
(similarly for $D_F$, $F_{D\mu\nu}$), and we have
\bea
D_{\mu}\phi&=&\pa_{\mu}\phi+iA^F_{\mu}\phi,\nn
F_{\mu\nu}^A&=&\pa_{\mu}A_{\nu}^A-\pa_{\nu}A_{\mu}^A-f^{ABC}
A_{\mu}^BA_{\nu}^C,
\label{Dmudef}
\eea
with similar definitions for $D_{\mu}\psi$, $D_{\mu}\lambda$.
If one decomposes $U(N)$ as
$SU(N)\otimes U(1)$ then our convention is that $\phi^a$ (for example) 
are the $SU(N)$
components and $\phi^0$ the $U(1)$ component.
Of course then $f^{ABC}=0$
unless all indices are $SU(N)$. 
We note that $d^{ab0}=\sqrt{\frak2N}\delta^{ab}$, $d^{000}=\sqrt{\frak2N}$.
We also have
\be
e^{ab}=\frac{1}{g^2},\quad e^{00}=\frac{1}{g_0^2}, \quad e^{0a}=e^{a0}=0.
\label{etensor}
\ee
Compared with our previous work such as Ref.~\cite{jjwc}, we have absorbed a 
factor of $g$ into our definitions of the fields in the gauge multiplet.
We have omitted terms which are $\Ncal=\frak12$ supersymmetric on
their own (such as terms involving only $\phibar$, $\lambdabar$ 
and/or $F$), which will have no relevance for our current discussion. 
They were 
considered in full in Refs.~\cite{penroma}; and indeed we
included them ourselves in Refs.~\cite{jjwa}, \cite{jjwb}.
We have, however, included some additional
sets of terms (those multiplied by $\kappa_{1-5}$)
which are required for renormalisability of the theory. Each of these sets
of terms is separately $\Ncal=\frak12$ invariant. 

It is easy to show that Eq.~(\ref{Sadj})\ is invariant under
\bea
\delta A^A_{\mu}&=& -i\lambdabar^A\sigmabar_{\mu}\epsilon,\nn  
\delta \lambda^A_{\alpha}&=& i\epsilon_{\alpha}D^A+\left(\sigma^{\mu\nu}\epsilon
\right)_{\alpha}\left[F^A_{\mu\nu}
+\frak12iC_{\mu\nu}d^{ABC}\lambdabar^B\lambdabar^C\right],\quad
\delta\lambdabar^A_{\alphadot}=0,\nn
\delta D^A&=& -\epsilon\sigma^{\mu}D_{\mu}\lambdabar^A,\nn
\delta\phi&=& \sqrt2\epsilon\psi,\quad\delta\phibar=0,\nn
\delta\psi^{\alpha}&=& \sqrt2\epsilon^{\alpha} F,\quad
\delta\psibar_{\alphadot}=-i\sqrt2(D_{\mu}\phibar)
(\epsilon\sigma^{\mu})_{\alphadot},\nn
\delta F^A&=& 0,\nn
\delta \Fbar^A&=& -i\sqrt2D_{\mu}\psibar^A\sigmabar^{\mu}\epsilon
-2i(\phibar\epsilon\lambda^F)^A
+2C^{\mu\nu}D_{\mu}(\phibar^B\epsilon\sigma_{\nu}   
(\lambdabar^D)^{AB}).
\label{newsusy}
\eea
In Eq.~(\ref{Sadj}), $C^{\mu\nu}$ is related to the non-anti-commutativity 
parameter $C^{\alpha\beta}$ by  
\be
C^{\mu\nu}=C^{\alpha\beta}\epsilon_{\beta\gamma}
\sigma^{\mu\nu}_{\alpha}{}^{\gamma},
\label{Cmunu}
\ee
where 
\bea
\sigma^{\mu\nu}&=&\frak14(\sigma^{\mu}\sigmabar^{\nu}-
\sigma^{\nu}\sigmabar^{\mu}),\nn
\sigmabar^{\mu\nu}&=&\frak14(\sigmabar^{\mu}\sigma^{\nu}-
\sigmabar^{\nu}\sigma^{\mu}).
\label{sigmunu}
\eea
Our conventions are in accord with \cite{seib}; in particular, 
\be
\sigma^{\mu}\sigmabar^{\nu}=-\eta^{\mu\nu}+2\sigma^{\mu\nu}.
\label{sigid}
\ee
Properties of $C$ which follow from
Eq.~(\ref{Cmunu})\ are  
\bea
C^{\alpha\beta}&&=\frak12\epsilon^{\alpha\gamma}
\left(\sigma^{\mu\nu}\right)_\gamma{}^{\beta}C_{\mu\nu},
\nn
C^{\mu\nu}\sigma_{\nu\alpha\betadot}&&=C_{\alpha}{}^{\gamma}
\sigma^{\mu}{}_{\gamma\betadot},\nn
C^{\mu\nu}\sigmabar_{\nu}^{\alphadot\beta}&&=-C^{\beta}{}_{\gamma}
\sigmabar^{\mu\alphadot\gamma}.
\label{cprop}
\eea 
We use the standard gauge-fixing term 
\be
S_{\rm{gf}}={1\over{2\alpha}}\int d^4x e^{AB}(\pa.A)^A(\pa.A)^B
\label{gafix}
\ee 
with its associated
ghost terms.  The vector propagator is given by  
\be
\Delta^{AB}_{V\mu\nu}=-{1\over{p^2}}\left(\eta_{\mu\nu}
+(\alpha-1){p_{\mu}p_{\nu}\over{p^2}}\right)\left(e^{-1}\right)^{AB}.
\label{gprop}
\ee
The scalar propagator is 
\be
\Delta_{\phi}^{AB}=-\frac{1}{p^2}P^{AB}
\label{sprop}
\ee
where
\be
P^{ab}=\delta^{ab},\quad P^{00}=\frac{1}{\kappa}, \quad P^{0a}=P^{a0}=0,
\ee
the fermion propagator is  
\be
\Delta^{AB}_{\psi\alpha\alphadot}=
{p_\mu\sigma^{\mu}_{\alpha\alphadot}\over{p^2}}P^{AB},
\label{fprop}
\ee
where the momentum enters at the end of the propagator with the undotted 
index,
and the auxiliary propagator is
\be
\Delta_F^{AB}=P^{AB}.
\label{aprop}
\ee

\section{Renormalisation}
The bare action will
be given as usual by replacing fields and couplings by their bare versions,
shortly to be given more explicitly.
Note that in the $\Ncal=\frak12$ supersymmetric case, fields and their
conjugates may renormalise differently. 
We found in Refs.~\cite{jjwa}, \cite{jjwb} 
that non-linear renormalisations of $\lambda$
and $\Fbar$ were required; and in a subsequent
paper\cite{jjwd} we pointed out that non-linear
renormalisations of $F$, $\Fbar$ are required even in ordinary $\Ncal=1$
supersymmetric gauge theory when working in the uneliminated formalism.
The renormalisations of the remaining fields and couplings are linear as 
usual (except for $\kappa$, $\kappa_{1-5}$, see later) and given by
(in the case of the $SU(N)$ fields)
\bea 
\lambdabar^a_B=Z_{\lambda}^{\frak12}\lambdabar^a,
\quad
A^{a}_{\mu B}=Z_A^{\frak12}A^{a}_{\mu}, 
&\quad& \phi^a_B=Z_{\phi}^{\frak12}\phi^a,\quad
\psi^a_B=Z_{\psi}^{\frak12}\psi^a,\nn
\phibar^a_B=Z_{\phi}^{\frak12}\phibar^a,
\quad \psibar^a_B=Z_{\psi}^{\frak12}\psibar^a, &\quad&
g_B=Z_gg,\quad y_B=Z_yy,\nn  
C_B^{\mu\nu}=Z_CC^{\mu\nu}, \quad (\kappa-1)_B&=&Z_{\kappa}(\kappa-1),\quad 
\kappa_{1-5B}=Z_{1-5}.
\label{bare}
\eea
The corresponding $U(1)$ gauge multiplet fields 
$\lambdabar^0$ etc are unrenormalised;
so is $g_0$. The renormalisation constants for the $U(1)$ chiral fields
will be denoted $Z_{\phi^0}$ etc and discussed later.
In Eq.~(\ref{bare}), $Z_{1-5}$ are divergent
contributions; in other words we have set the renormalised couplings
$\kappa_{1-5}$ to zero for simplicity.   
The anomalous dimensions $Z_{\lambda}$ etc, and the renormalisation 
constants for the couplings $g$, $y$,
$C$ and $(\kappa-1)$, start with tree-level values of 1. (The slightly
non-standard definition of $Z_{\kappa}$ is once again to make our results
correspond more closely with those of Ref.~\cite{penroma}.) 
The anomalous dimensions for the gauge-multiplet fields
and hence the gauge $\beta$-functions are the same as in the standard 
$\Ncal=1$ theory. 
Since our gauge-fixing term in Eq.~(\ref{gafix})\ does not preserve 
supersymmetry, the anomalous dimensions $Z_A$ and $Z_{\lambda}$
for $A^a_{\mu}$ and $\lambda^a$ are
different (and moreover gauge-parameter dependent), as are those ($Z_{\phi}$
and $Z_{\psi}$) for $\phi^a$ and $\psi^a$. Moreover, neither 
$Z_{\phi}$ nor $Z_{\psi}$ coincide with  
$Z_{\Phi}$, the chiral superfield renormalisation constant. 

We have assigned the same coupling $y$ to all the three-point interactions;
for instance, both $d^{abc}\phi^a\phi^b\phi^c$ and 
$d^{0bc}\phi^0\phi^b\phi^c$. This is by no means guaranteed a priori. From
the non-renormalisation theorem, one expects 
\be
Z_y=Z_{\Phi}^{-\frac32}
\label{nonren}
\ee
and so consistency requires $Z_{\Phi}$ and $Z_{\Phi^0}$ to be equal.
This is arranged by a judicious choice of $Z_{\kappa}$ (a change 
in $Z_{\kappa}$ alters $Z_{\Phi}$ while leaving $Z_{\Phi^0}$ unchanged).  

At one loop we find, writing $Z^{(n)}$ for the $n$-loop contribution to $Z$, 
\bea
Z^{(1)}_{\phi}&=&[-N'y\ybar+2g^2(1-\alpha)N]L,\nn
Z^{(1)}_{\psi}&=&[-N'y\ybar-2g^2(1+\alpha)N]L,\nn
Z^{(1)}_F&=&-N'y\ybar L,\nn
Z^{(1)}_y&=&-\frac32Z_{\Phi}^{(1)},\nn
Z_{\Phi}^{(1)}&=&[-N'y\ybar
+4g^2N]L,\nn
Z_g&=&1-2g^2NL,
\label{zchi}
\eea
where (using dimensional regularisation with $d=4-\epsilon$)
\be
L={1\over{16\pi^2\epsilon}}
\ee
and 
\be
N'=N+\frac{4}{N\kappa}(1-\kappa).
\label{Ndef}
\ee
The remaining renormalisation constants will not be required but 
the one-loop results can be found in Ref.~\cite{jjpb}. 
The
difference between $Z_{\Phi}$ and $Z_{\phi}$, $Z_{\psi}$ is due solely   
to the choice of a non-supersymmetric gauge; the gauge-independent terms are
the same, and since there are no gauge interactions for the $U(1)$ fields
anyway, we have 
\be
Z_{\Phi^0}=Z_{\phi^0}=Z_{\psi^0}.
\ee
We now choose
\bea
Z^{(1)}_{\kappa}&=&-{4g^2N\kappa\over{\kappa-1}}+
{y\ybar N(\kappa-2)\over{\kappa-1}}
-{2y\ybar(2\kappa^2-\kappa-1)\over{N\kappa^2}}
\label{zlam}
\eea
which guarantees that $Z_{\Phi^0}$ and $Z_{\Phi}$ match at one loop.

We have now dealt with the majority of the renormalisations of fields and 
couplings.  
The remaining non-linear renormalisations of $\lambda$, $F$ and $\Fbar$ are
largely determined in order to cancel $C$-dependent 
divergences; though as we have emphasised, 
a non-linear renormalisation of $F$ and $\Fbar$ is required
in the usual $\Ncal=1$ ($C=0$) case. The precise forms of these 
non-linear renormalisations
are not required for our computation, as we shall explain; and will therefore
be omitted, though once again they can be found 
(at one loop) in Ref.~\cite{jjpb}. We then found in Ref.~\cite{jjpb} that
$C$ is unrenormalised at one loop, i.e. $Z_C^{(1)}=0$.
Our main interest is in determining whether the $C$ parameter remains 
unrenormalised at the two loop level. To this end,
the simplest approach appeared to be to focus on the 
$\ybar C^{\mu\nu}f^{abc}\pa_{\mu}\phibar^a \pa_{\nu}\phibar^b\phibar^c$
term. There are two reasons for this. Firstly, fermion calculations 
frequently produce the quantities 
\be
\epsilon^{\mu\nu\rho\sigma}\sigma_{\rho\sigma},
\quad \epsilon^{\mu\nu\rho\sigma}\sigmabar_{\rho\sigma},
\ee
where $\epsilon^{\mu\nu\rho\sigma}$ is the four-dimensional alternating 
symbol and $\sigma_{\rho\sigma}$ , 
$\sigmabar_{\rho\sigma}$ are defined in Eq.~(\ref{sigmunu}). 
In exactly four dimensions we have
\be
\epsilon^{\mu\nu\rho\sigma}\sigma_{\rho\sigma}=2\sigma^{\mu\nu},
\quad \epsilon^{\mu\nu\rho\sigma}\sigmabar_{\rho\sigma}=-2\sigmabar^{\mu\nu}
\label{epsid}
\ee
(i.e. $\sigma_{\rho\sigma}$, $\sigmabar_{\rho\sigma}$ are selfdual
and anti-self-dual respectively)
but it is not clear if these identities remain true away from four dimensions
and can therefore be used in the context of dimensional regularisation.
Choosing a purely bosonic interaction seems likely to 
reduce the numbers of appearances of these
quantities, and in fact we shall find it never appears in 
our calculation (though we do meet the quantity
$\epsilon^{\mu\nu\rho\sigma}C_{\rho\sigma}$, which we shall discuss shortly).  
Secondly, this interaction contains no auxiliary field $F$
and hence is unaffected by any non-linear renormalisation 
of the auxiliary field $\Fbar$ which would otherwise also need to be 
determined in order to fix the value of $Z_C$. 

\begin{figure}[H]
\includegraphics{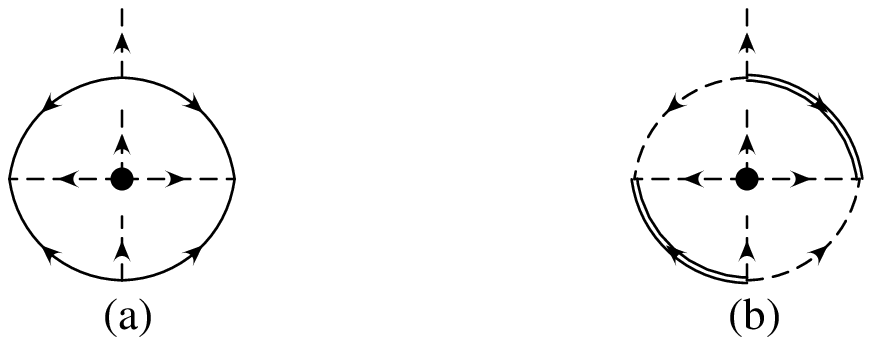}
\caption{Two-loop $(y\ybar)^2$ graphs} \label{fig1}
\end{figure}

\begin{figure}[H] 
\includegraphics{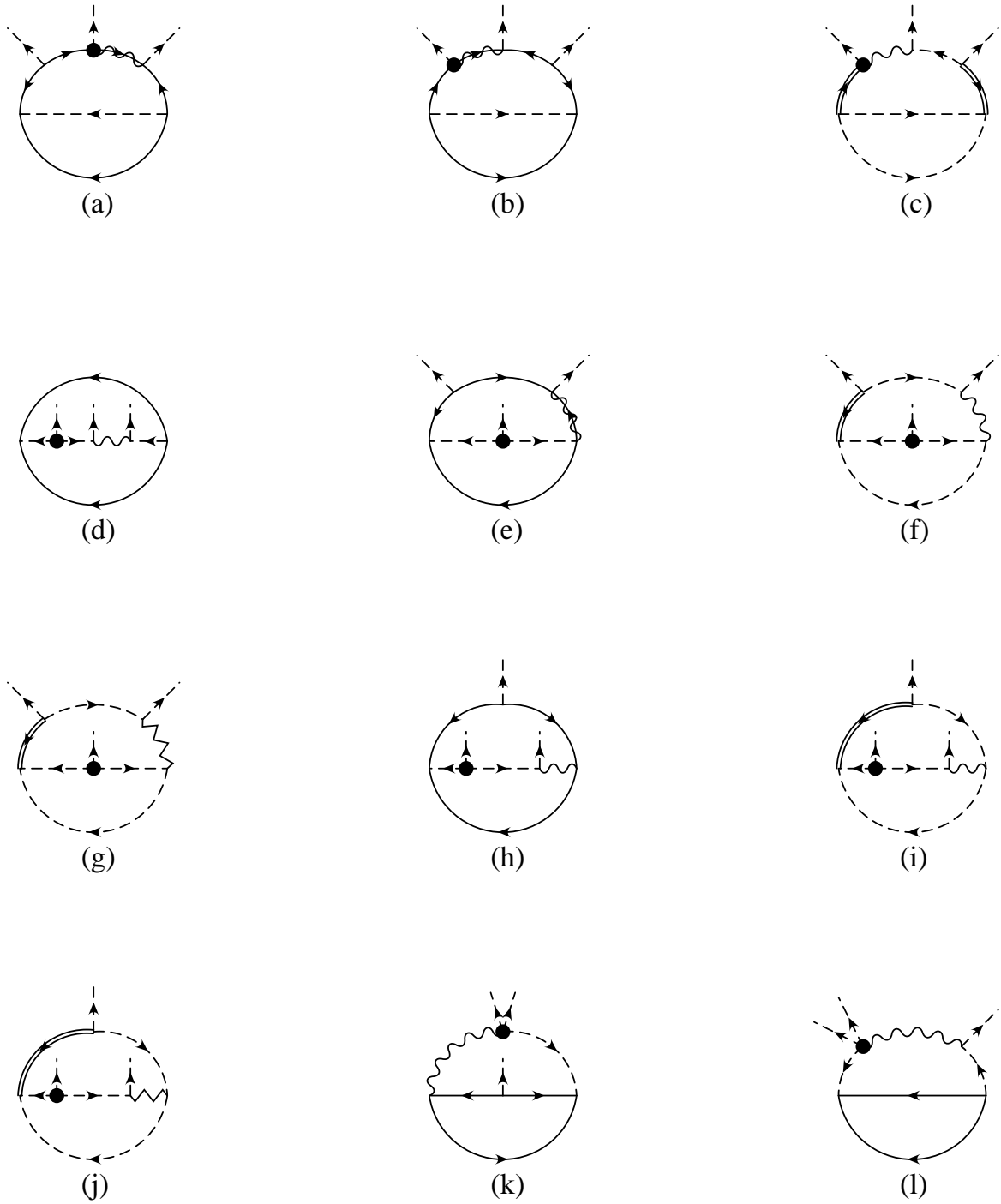}
\caption{Two-loop $g^2y\ybar$ graphs} \label{fig2}
\end{figure}

\begin{figure}[H]
\includegraphics{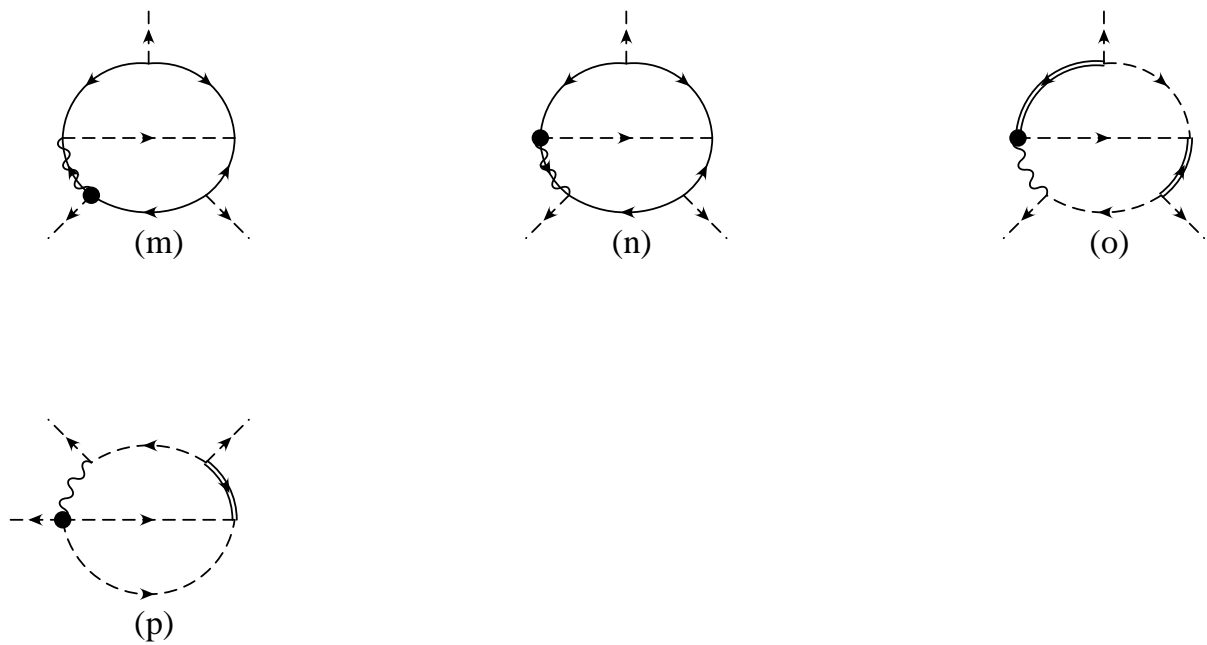}
\caption{Two-loop $g^2y\ybar$ graphs (continued)} \label{fig3}
\end{figure}

\begin{figure}[H]
\includegraphics{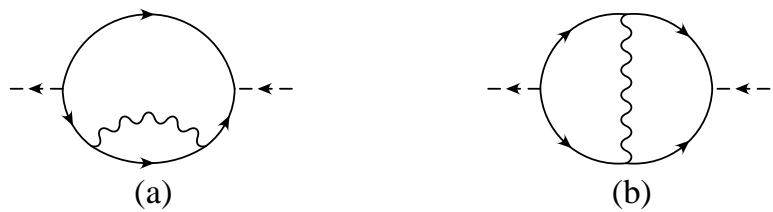}
\caption{Graphs for $Z_{\phi}$ contributing to DREG/DRED difference}\label{fig4}
\end{figure}

As we mentioned earlier, we shall only consider the Yukawa-dependent terms 
in the two-loop renormalisation constant for $C$. The main reason for 
setting aside the remaining 
graphs (which would contribute $g^4$ terms at this loop order) is their sheer 
number, namely around a hundred; but there are other technical reasons 
which we shall discuss in due course.  
The two-loop diagrams which contribute to the $(y\ybar)^2$ terms in the  
renormalisation of 
$\ybar C^{\mu\nu}f^{abc}\pa_{\mu}\phibar^a \pa_{\nu}\phibar^b\phibar^c$ 
are depicted in Fig.~\ref{fig1} and those which contribute to the 
$g^2y\ybar$ terms
are shown in Figs.~\ref{fig2}, \ref{fig3}. 
The diagrams in Fig.~\ref{fig1} give a vanishing contribution on
grounds of symmetry (in fact we have omitted several diagrams from 
Figs.~\ref{fig2}, \ref{fig3} 
which give vanishing contributions for similar reasons, in 
addition to diagrams which give no logarithmic divergences after subtraction
of divergent subdiagrams). For instance, diagrams with two 
$\phibar$ (and not $\pa\phibar$) lines emerging from the same vertex 
or connected by an auxiliary chiral propagator 
are zero by symmetry. The divergent contributions from the diagrams (a)-(p) 
in Fig.~\ref{fig2}, \ref{fig3} are denoted by $G_1-G_{16}$ respectively and
listed below (we perform subtractions of subdivergences on a 
diagram-by-diagram basis, so that individual results are purely local):
\bea
G_1&=&2L^2N_1(1-\frak34\epsilon),\nn
G_2&=&2L^2\left[N^2+2\left(\frac{6}{\kappa}-4\right)+8K_1\right]
(1-\frak{1}{2}\epsilon),\nn
G_3&=&\frak{1}{2}L^2
\left[N^2+2\left(\frac{6}{\kappa}-4\right)+8K_1\right]\epsilon,\nn
G_4&=&\alpha L^2N_1X,\nn
G_5 &=&-L^2N_2X,\nn
G_{6}&=&\frak34L^2N_2(1-\frak{7}{12}\epsilon),\nn
G_{7}&=&\frak14L^2N_2\epsilon,\nn
G_8&=&-\alpha L^2 N_1X,\nn
G_{9}&=&0,\nn
G_{10}&=&\frak14L^2N_1\epsilon,\nn
G_{11}&=&\frak12L^2N_1[3+\frak12(\alpha-\frak12)\epsilon],\nn
G_{12}&=&-\frak12L^2N_1[3+\frak12(\alpha-\frak{7}{2})\epsilon],\nn
G_{13}&=&-2L^2N_1X,\nn
G_{14}&=&-L^2\left[N^2+4\left(\frac{4}{\kappa}-3\right)+16K_1\right]X,\nn
G_{15}&=&\frak12L^2\left[N^2+4\left(\frac{4}{\kappa}-3\right)+16K_1\right]
\epsilon,\nn
G_{16}&=&-\frak34L^2N_2(1-\frak{17}{36}\epsilon)\nn
&-&\frak23L^2\left[N^2+\left(\frac{14}{\kappa}-10\right)+12K_1\right]
\epsilon,
\eea
where 
\be
X=1-\frak14\epsilon
\ee
and
\bea
N_1&=&NN'=N^2+4\left(\frac{1}{\kappa}-1\right),\nn
N_2&=&N^2+4\left(\frac{2}{\kappa}-1\right),\nn
K_1&=&\frac{1}{N^2}\left(2-\frac{3}{\kappa}+\frac{1}{\kappa^3}\right),
\eea
with $N'$ as defined in Eq.~(\ref{Ndef}). Some group identities used in 
deriving these results are listed in the Appendix.
Note that the ``deformed'' vertex
in diagram (p) contains contributions from both terms in the 9th line
of Eq.~(\ref{Sadj}).   
In the case of the majority of diagrams, the only property we have assumed
for $\epsilon^{\mu\nu\rho\sigma}$ and $C^{\mu\nu}$ is that they are (totally)
antisymmetric tensors. However, in deriving the result for diagram (n) 
we have also assumed that the identity
\be
\epsilon^{\mu\nu\rho\sigma}C_{\rho\sigma}=2C^{\mu\nu}
\label{pres}
\ee
(which is valid in four dimensions--i.e. $C_{\mu\nu}$ is self-dual) remains 
true in $d=4-\epsilon$ dimensions. This seems to be a natural requirement and 
somewhat in the spirit of dimensional reduction; in any case, we shall discuss
this choice in more detail later. 
We then readily find
\be
\sum_1^{16}G_i=0,
\ee
which implies 
\be
\left(Z_CZ_yZ_{\phi}^{\frak32}\right)^{(2)}_{\rm{pole},y-\rm{dependent}}=0.
\label{baretwo}
\ee
In order to determine $Z_C$ we need to know $Z_y$ and $Z_{\phi}$ through
two loops.  
The two-loop result for $Z_{\phi}^{(2)}$ is given in Ref.~\cite{mach} 
and translates (in our conventions) to
\bea
Z_{\phi,\rm{DREG}}^{(2)}&=&L^2(y\ybar)^2\left[-N_1^2
+\frac{2}{\kappa N^2}(N^2-4)+\frac{4}{\kappa^2N^2}(N^2-2)+\frac{4}{\kappa^4N^2}
\right](1-\frak34\epsilon)\nn
&&+L^2y\ybar g^2\left[2N_1(2+\alpha)-\frac{8}{\kappa}
-\left(N^2-4+\frac{2}{\kappa}\right)\epsilon\right]+\ldots.
\label{dreg} 
\eea
where the ellipsis indicates $g^4$ terms which we shall not require.
This result is computed (as the notation indicates) using dimensional 
regularisation (DREG) (where the number of gauge fields becomes $d$ in
$d$ dimensions), while for a supersymmetric calculation we should be using
dimensional reduction (DRED) (where the number of gauge fields is maintained
as exactly four, even in $d$ dimensions).
The difference between our $Z_{\phi}^{(2)}$ and $Z_{\phi,\rm{DREG}}^{(2)}$
resides (as far as the $y$-dependent terms are concerned) in the two diagrams
in Fig.~\ref{fig4} which contribute to $Z_{\phi}$--specifically in the two
$\sigma$ matrices contracted by the gauge propagator. We find
\be 
Z_{\phi}^{(2)}-Z_{\phi,\rm{DREG}}^{(2)} = \frac{2}{\kappa}L^2g^2y\ybar\epsilon.
\label{diff}
\ee
This conclusion may be confirmed using the results of Ref.~\cite{JO}. This
reference presented results for $Z_{\phi}$ at two loops computed using 
DRED, together with expressions for the differences between $\beta$ 
functions computed using the two schemes. However we cannot use these DRED 
results directly because they are presented in the Feynman background 
gauge and we require results for a general conventional gauge. Luckily, the
difference between DRED and DREG for the $g^2y\ybar$ terms is 
gauge-independent.
Combining Eqs.~(\ref{dreg}), (\ref{diff}) we have
\bea
Z_{\phi}^{(2)}&=&L^2(y\ybar)^2\left[-N_1^2
+\frac{2}{\kappa N^2}(N^2-4)+\frac{4}{\kappa^2N^2}(N^2-2)+\frac{4}{\kappa^4N^2}
\right](1-\frak34\epsilon)\nn
&&+L^2y\ybar g^2\left[2N_1(2+\alpha)-\frac{8}{\kappa}
-(N^2-4)\epsilon\right].
\label{ztwoa}
\eea
The result for $Z_{\Phi}^{(2)}$ may be extracted from Ref.~\cite{west}
\bea
Z_{\Phi}^{(2)}&=&L^2(y\ybar)^2\left[-N_1^2
+\frac{2}{\kappa N^2}(N^2-4)+\frac{4}{\kappa^2N^2}(N^2-2)+\frac{4}{\kappa^4N^2}
\right](1-\frak34\epsilon)\nn
&&+L^2y\ybar g^2(N^2-4)(2-\epsilon).
\label{ztwob}
\eea
As a check the double poles may also be obtained from the one-loop results 
using 
\be
Z_{\phi\rm{double pole}}^{(2)}=\frak12\left\{\left(Z_{\phi}^{(1)}\right)^2+
\left[Z_{y}^{(1)}y\frac{\pa}{\pa y}+Z_{\ybar}^{(1)}\ybar\frac{\pa}{\pa \ybar}
+Z^{(1)}_gg\frac{\pa}{\pa g}+Z^{(1)}_{\kappa}(\kappa-1)
\frac{\pa}{\pa \kappa}\right]
Z_{\phi}^{(1)}\right\}
\ee
with a similar result for $Z_{\Phi}$.

Using Eqs.~(\ref{nonren}), (\ref{zchi}), (\ref{ztwoa}), (\ref{ztwob})
we can conclude that
\be
Z_C=O(g^4)
\ee
up to two loops. 
\section{Conclusions} 
We have found that $Z_C$ is unrenormalised through two loops, as far as the 
Yukawa-dependent contributions are concerned; this seems a strong indication
that $Z_C$ is completely unrenormalised at this order. Nevertheless it would
be reassuring to compute $Z_C$ in full, so  
we shall now discuss further
our choice of prescription Eq.~(\ref{pres}) for $C^{\mu\nu}$ and also the
feasibility of completing the calculation by including the remaining $g^4$ 
terms. 

In four dimensions we have the identity 
\bea
\sigma^{\mu}\sigmabar^{\nu}\sigma^{\rho}
&=&-\eta^{\mu\nu}\sigmabar^{\rho}-\eta^{\nu\rho}\sigmabar^{\mu}
+\eta^{\mu\rho}\sigmabar^{\nu}+\epsilon^{\mu\nu\rho\sigma}\sigmabar_{\sigma},\nn
\sigmabar^{\mu}\sigma^{\nu}\sigmabar^{\rho}
&=&-\eta^{\mu\nu}\sigma^{\rho}-\eta^{\nu\rho}\sigma^{\mu}
+\eta^{\mu\rho}\sigma^{\nu}-\epsilon^{\mu\nu\rho\sigma}\sigma_{\sigma}.
\label{epsida}
\eea
By contracting Eqs.~(\ref{epsida}) with $\sigmabar_{\rho}$, 
$\sigma_{\rho}$ it is easy to derive the identities  
\be
\epsilon^{\mu\nu\rho\sigma}\sigma_{\rho\sigma}=(d-2)\sigma^{\mu\nu},
\quad \epsilon^{\mu\nu\rho\sigma}\sigmabar_{\rho\sigma}=-(d-2)
\sigmabar^{\mu\nu},
\label{epsidb}
\ee
In order to reconcile Eqs.~(\ref{epsidb}), (\ref{pres}) it seems that we 
must abandon the identity Eq.~(\ref{Cmunu}) (or at least modify it away from
two dimensions). The reason that we have not had to confront this issue so far
in our calculation might be that to leading order one can prove the 
invariance of the chiral part of Eq.~(\ref{Sadj}) under Eq.~(\ref{newsusy})
without assuming anything about $C^{\mu\nu}$ other than its antisymmetry; in
particular the relation Eq.~(\ref{Cmunu}) between $C^{\mu\nu}$ and
$\sigma^{\mu\nu}$ is not required. Of course an alternative would be to impose
a different prescription to Eq.~(\ref{pres}), for instance one analogous to
Eq.~(\ref{epsidb}); however this would change the simple pole term
for Fig.~\ref{fig2}(k) and would introduce (amongst other terms) a 
$\frac{1}{\kappa^3}$ term into $Z_C$. On the other hand, it can be seen that 
there is no diagram contributing to the renormalisation of (for instance)
$C^{\mu\nu}d^{ABC}e^{AD}F^D_{\mu\nu}\lambdabar^B\lambdabar^C$ in 
Eq.~(\ref{Sadj}) which could contain $\frac{1}{\kappa^3}$ dependence; nor
does it appear in $Z_{\phi}$ or $Z_y$. Therefore it cannot occur in $Z_C$.

Clearly we would like to see whether $Z_C$ really does vanish completely through
two loops by examining the remaining $g^4$-type diagrams. However the complete
proof of the $\Ncal=\frak12$ invariance of Eq.~(\ref{Sadj}) requires use 
of Eq.~(\ref{cprop}) which in turn depends on Eq.~(\ref{Cmunu}). It therefore
seems that it may be difficult to find a consistent definition for 
$C^{\mu\nu}$ which will maintain the complete invariance of Eq.~(\ref{Sadj})
in general $d$ dimensions, and this means that we may not expect to obtain an
unambiguous answer for the $g^4$ contribution to $Z_C$ within dimensional
regularisation. Indeed, although it does not appear that the potentially
ambiguous quantity
$\epsilon^{\mu\nu\rho\sigma}\sigma_{\rho\sigma}$ arises in any of the $g^4$
graphs, we do find ourselves obliged to simplify contractions of the form
$\epsilon^{\mu\nu\rho\sigma}\epsilon_{\rho\sigma\alpha\beta}$. In exactly four
dimensions we have
\be
\epsilon^{\mu\nu\rho\sigma}\epsilon_{\kappa\lambda\alpha\beta}
=4!\delta^{\mu}{}_{[\kappa}\delta^{\nu}{}_{\lambda}
\delta^{\rho}{}_{\alpha}\delta^{\sigma}{}_{\beta]}.
\ee
If we assume that this result remains true away from four dimensions then
we obtain
\be
\epsilon^{\mu\nu\rho\sigma}\epsilon_{\rho\sigma\alpha\beta}
=2(d-2)(d-3)\delta^{\mu}{}_{[\alpha}\delta^{\nu}{}_{\beta]},
\ee
but once again the consistency with Eqs.~(\ref{epsidb}), (\ref{pres})
is a moot point. 

A possible alternative approach to the two-loop calculation could be the use of 
differential regularisation\cite{freed} which enables one to work in exactly 
four dimensions. However if one accepts that the results we have so far
obtained are a strong indication of the non-renormalisation of $Z_C$ at two 
loops, then a more fruitful approach may be to seek a general 
proof of the result to all orders. 
One may speculate that the non-renormalisation of the
non-anticommutativity parameter may follow from some kind of analogue of the
Slavnov-Taylor identities. This might be somewhat difficult to see in this
component formulation but might be more transparent in the superspace 
formalism combined with the background field formalism\cite{penrom,grisb,
penroma,penromb}; where the result would be more comparable to a simple Ward 
identity, due to the manifest supersymmetry and gauge 
invariance in this case\cite{penati}.

\bigskip

\bigskip

\noindent 
{\large{\bf Acknowledgements\\}}
One of us (RP) was supported by an STFC studentship. IJ is grateful 
for a useful discussion with Silvia Penati.

\section{Appendix}
Identities for $SU(N)$ useful for simplifying the divergent contributions
are\cite{azcar}
\bea
\tr[\Dtil^a\Dtil^b]={N^2-4\over{N}}\delta^{ab},&\quad&
\tr[\Dtil^a\Dtil^b\Dtil^c]=\frac{N^2-12}{2N}d^{abc},\nn
\tr[\Ftil^a\Ftil^b\Dtil^c]=\frac{N}{2}d^{abc},&\quad&
\tr[\Ftil^a\Dtil^b\Dtil^c]=i\frac{N^2-4}{2N}f^{abc},\nn
\tr[\Dtil^a\Ftil^b\Dtil^d\Dtil^c\Dtil^d]&=&i\frac{(N^2-12)(N^2-4)}{4N^2}
f^{abc},\nn
\tr[\Dtil^a\Dtil^b\Dtil^d\Dtil^c\Ftil^d]&=&i\frac{(N^2-4)^2}{4N^2}f^{abc},\nn
\tr[\Dtil^a\Ftil^b\Ftil^d\Ftil^c\Dtil^d]=
\tr[\Ftil^a\Ftil^b\Ftil^d\Dtil^c\Dtil^d]&=&-i\frac{(N^2-4)^2}{4N^2}f^{abc}.
\eea
\vfill
\eject

\end{document}